\documentclass[aps,prl,reprint]{revtex4-1}
\usepackage{amsmath}
\usepackage{amsfonts}
\usepackage{amssymb}
\usepackage{mathtools}
\usepackage{graphicx}
\usepackage{tabularx}
\usepackage[caption=false,position=b]{subfig}
\usepackage[colorlinks]{hyperref}
\usepackage[capitalize]{cleveref}

\usepackage{yypreamble}

\newcommand{\vm}{\vec{m}}
\newcommand{\vn}{\vec{n}}

\newcommand{\pardash}[1]{\section*{#1}}

\begin{document}

\title{An algorithm for tailoring a quadratic lattice with a local squeezed reservoir to stabilize generic chiral states with non-local entanglement}
\author{Yariv Yanay}
\email{yariv@lps.umd.edu}
\affiliation{Laboratory for Physical Sciences, 8050 Greenmead Dr., College Park, MD 20740}

\date{\today}

\begin{abstract}
We demonstrate a new approach to the generation of custom entangled many-body states through reservoir engineering, using the symmetry properties of bosonic lattice systems coupled to a local squeezed reservoir \cite{Yanay2018}.
We outline an algorithm where, beginning with a desired set of squeezing correlations, one uses the symmetry to constrain the Hamiltonian and find a lattice configuration which stabilizes a pure steady state realizing these correlations.
We demonstrate how to use this process to stabilize two unique pure states with non-local correlations that could be useful for quantum information applications.
First, we show how drive a square lattice into a product state of entangled quadruplets of sites.
Second, using a bisected system, we generate a steady state where local measurements in one half of the lattice herald a pure delocalized state in the second half.
\end{abstract}

\maketitle

The generation of entangled non-classical states for quantum computation and other quantum information applications is a subject of ongoing interest and research. 
One method of tackling this challenge, known as reservoir engineering \cite{Poyatos1996,Plenio2002}, is to add a dissipative bath to a quantum system, carefully coupled so that the overall steady state is such a quantum state of interest. Reservoir engineering has seen growing theoretical and experimental exploration, from the stabilization of systems with few degrees of freedom \cite{Krauter2011,Murch2012,Lin2013,Shankar2013,Leghtas2015,Wollman2015}, through many mode systems with system-wide dissipation \cite{Diehl2008,Cho2011,Koga2012,Ikeda2013,Quijandria2013}, to, more recently, the preparation of many-mode states through systems coupled to a single local dissipative bath \cite{Zippilli2015,Ma2016,Ma2017,Ma2017a}.

Recently, we have shown that an entire class of bosonic lattice systems can be stabilized in this manner \cite{Yanay2018}. The existence of a ``generalized chiral symmetry'' compatible with a single, local dissipation source, implies that the steady state is a non-trivial, often highly non-local, pure squeezed state. We also proved that the chiral symmetry condition, formulated at the Hamiltonian matrix level, dictates the full form and correlations of the steady state.

In this letter, we show how this symmetry can be used as a powerful tool of reservoir engineering in the most straightforward sense: given a desired pattern of squeezing correlations, we propose an algorithm to engineer a quadratic lattice stabilizing a pure steady state realizing them. We provide two examples of this process, outlining the route from correlation matrix to lattice design: a two dimensional (2D) grid realizing a four-fold correlation pattern, and a heralding system where the occupation of momentum states in one sublattice is entangled with that of number states in a second.

\section{Model and Algorithm}
We begin by considering a $d$-dimensional bosonic lattice of $N$ sites, described by a generic particle conserving quadratic Hamiltonian,
\begin{equation}\begin{split}
\hat \H & =  \vec{\hat a}\dg\cdot H \cdot \vec{\hat a} = \sum_{\mathclap{\vm, \vn}}H_{\vm,\vn}\hat a_{\vm}\dg\hat a_{\vn}.
\label{eq:HS}
\end{split}\end{equation}
Here, summation is over all sites, labeled by $d$-dimensional vectors  $\vm, \vn$, and $\hat a_{\vn}$ ($\hat a_{\vn}\dg$) is the annihilation (creation) operator for a boson on site $\vn$. The Hamiltonian matrix $H$ consists of on-site potentials $H_{\vn,\vn} = V_{\vn}$ and hopping elements $H_{\vm,\vn} = J_{\vm,\vn}$. 
We do not assume any symmetry or translational invariance, and allow hopping between any two sites.

We take a single ``drain'' site, $\vn_{0}$, to be linearly coupled at strength $\Gamma$ to a squeezed zero-temperature Markovian reservoir. The system's evolution is given by \cite{Gardiner2004}
\begin{subequations}\begin{gather}
\dot{\hat \rho} = i\br{\hat \rho, \hat{ \mathcal H}} 
	+ \Gamma\p{\hat a\pr_{\vn_{0}}\hat \rho \hat a_{\vn_{0}}^{\prime\dagger} - \half\acom{\hat a_{\vn_{0}}^{\prime\dagger}\hat a_{\vn_{0}}\pr,\hat \rho}},
\\ \hat a_{\vn_{0}}\pr = \cosh r \hat a_{\vn_{0}} - e^{i\phi}\sinh r \hat a_{\vn_{0}}\dg
\end{gather}\end{subequations}
where $r,\phi$ are the squeezing parameter and angle, respectively \cite{Drummond2004}.
This Hamiltonian, along with the squeezed reservoir, can be realized experimentally by a range of Bosonic systems, including coupled arrays of superconducting cavities \cite{Fitzpatrick2017,Owens2018} or mechanical oscillators \cite{Ludwig2013}.

We have previously shown \cite{Yanay2018} that in the absence of ``dark'' modes, i.e.~eigenmodes of $H$ with vanishing wavefunction at the drain, the system relaxes to a unique steady state.
Furthermore, given a symmetric, unitary $N\times N$ matrix $\gs$ such that
\begin{equation}
\gs^{T} = \gs, \qquad \gs\dg\cdot \gs = \mathbb{I}, \qquad \gs_{\vm,\vn_{0}} = \gd_{\vm,\vn_{0}},
\label{eq:chiralscond}
\end{equation}
$\gs$ is a ``generalized chiral symmetry'' of the system, i.e.
\begin{equation}
\gs\dg\cdot H \cdot \gs = -H^{*},
\label{eq:chiralreq}
\end{equation}
if and only if the steady state is the pure squeezed state
\begin{equation}\begin{split}
 \ket{\psi_{\rm ss}} \propto \exp\br{e^{i\phi}\tanh r \sum_{\vm,\vn} \gs_{\vm,\vn}\hat a_{\vm}\dg\hat a_{\vn}\dg}\vac.
\label{eq:GCSss}
\end{split}\end{equation}

Depending on the the nature of $\gs$, this steady state can contain a large amount of long-range entanglement, with strength depending on $r$. This is made manifest when one considers its anomalous correlations, given by $\bra{\psi_{\rm ss}}\hat a_{\vm}\hat a_{\vn}\ket{\psi_{\rm ss}} = \gs_{\vm,\vn} e^{i\phi}\cosh r \sinh r$. This entanglement can be used a resource in multiple ways: the state $\ket{\psi_{\rm ss}}$ is similar to the cluster states used in continuous-variable quantum computing \cite{Raussendorf2001,Gu2009,Menicucci2011}; squeezed light can be converted into entanglemd qubit states \cite{Kraus2004}, affording a resource for digital quantum computing; and finally, as we describe below, two-mode squeezed states are a natural fit for heralding systems.

We proceed by observing that the logic of the derivation of the symmetry condition can be followed in reverse, beginning with a desired set of correlations and finding a Hamiltonian that realizes them in its steady state. This leads directly to an algorithm for lattice engineering:
\begin{enumerate}
\item \emph{Choose a desired correlation matrix $\gs$, which satisfies the constraints of \cref{eq:chiralscond}.}

This is a conceptual state, driven by the desired application, subject to the unitarity and symmetry conditions. It may be inspired by previous chiral systems, as we do below for the four-fold symmetry.

\item \emph{Write down the the terms of the Hamiltonian matrix $H$ that are experimentally feasible or desirable.}

This is dictated by experimental constraints: for example, in a superconducting circuit, device topology may limit one to coupling nearest neighbor or next-nearest neighbor sites only; or long distance coupling may allowed but only in the form of a specific all-to-all coupling mediated by a cavity.

\item \emph{Obtain a set of constraints from \cref{eq:chiralreq}, and substitute terms in $H$ as necessary.}

This step is analytical in nature, involving the solution of a simple set of linear equations. 

\item \emph{Vary the remaining terms to ensure that there are no degeneracies or dark modes.}

This part can be done numerically for any experimentally realizable system, by calculating the eigenvectors and eigenvalues of the matrix, ${H\cdot  \vec\psi^{\p{i}} =\gep_{i} \vec\psi^{\p{i}}}$. These correspond to the eigenmode wavefunction and energies of the original Hamiltonian. Dark modes arise when $ \vec\psi^{\p{i}}_{\vn_{0}} = 0$, or when there is a spectrum degeneracy $\gve_{i} = \gve_{j}$, and so the robustness of the steady state is characterized by the mode most weakly coupled to the drain, $\min_{i}|\vec\psi^{\p{i}}_{\vn_{0}}|$, and by the point in the spectrum closest to degeneracy, $\min_{i,j}\abs{\gve_{i}-\gve_{j}}$. To minimize sensitivity to any imperfection one would seek a design which maximizes the value of both.

\end{enumerate}

We now apply this process in two cases of interest.

\begin{figure}[thbp] 
   \centering
   \includegraphics[width=0.75\columnwidth]{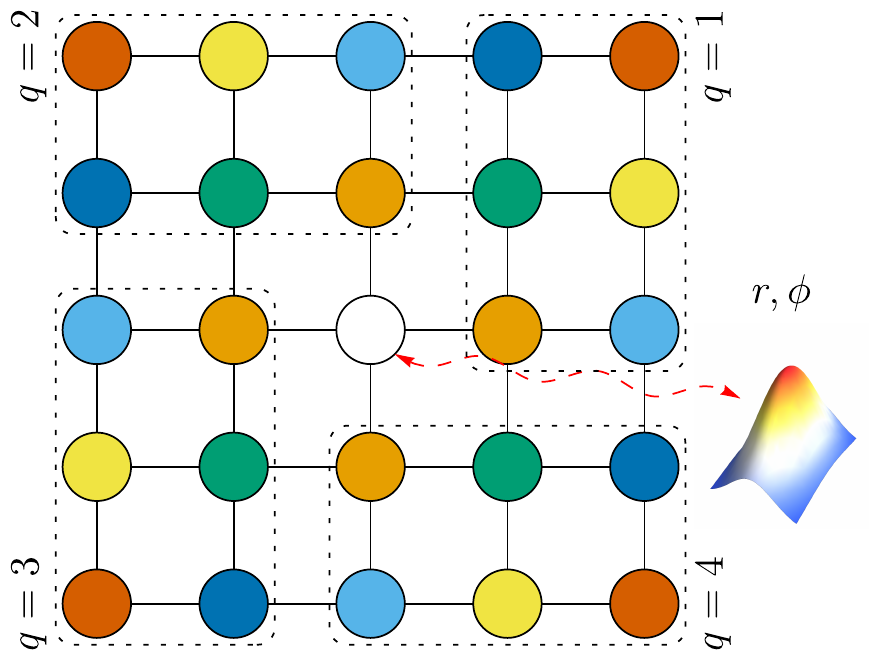}

   \caption{Four-fold correlation pattern in a 2D lattice. At the steady state, each quadruplet of sites of the same color, consisting of the four corners of a square centered at the origin, is entangled only amongst itself, with correlation pattern given by $\gs^{\p{\rm 4f}}_{\vec m,\vec n }$.
   The drain site is placed at the center, coupled to a squeezed bath with parameters $r,\phi$. The four quadrants of the lattice used in the notation of \cref{eq:nquad} are outlined as well.
   }
   \label{fig:4foldcorr}
\end{figure}

\pardash{Four-fold Entanglement}
\begin{figure*}[t] 
   \centering
   \hfill
   
   \subfloat[Plaquette flux]{\includegraphics[width=0.35\columnwidth]{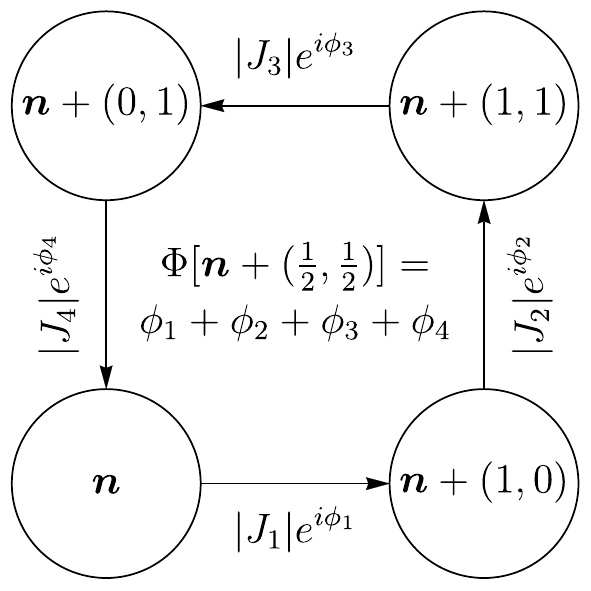} \label{fig:4foldflux}}
   \hfill
   \subfloat[Robustness to dark modes]{\parbox[b]{0.8\columnwidth}{
   	\includegraphics[width=0.75\columnwidth]{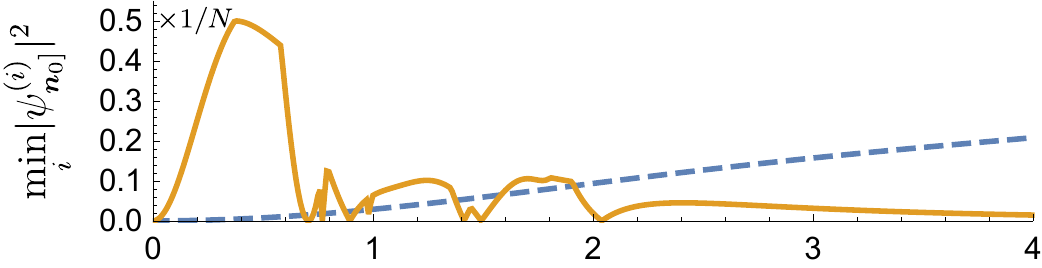}
   	\\ \vspace{5pt}\includegraphics[width=0.75\columnwidth]{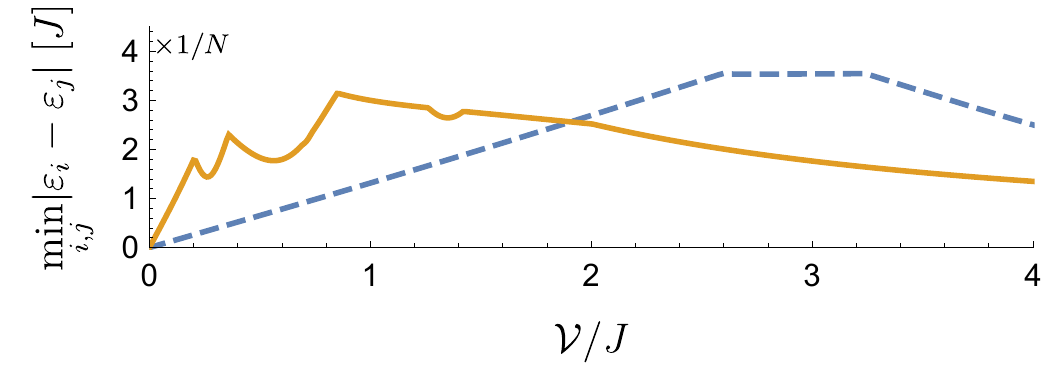}}
	\label{fig:4folddV}
   }
   \parbox[b]{0.35\columnwidth}{
   	 \includegraphics[width=0.33\columnwidth]{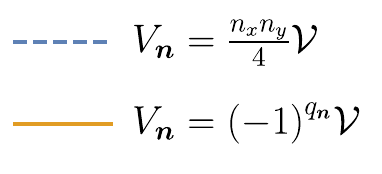}\hfill\hfill\hfill
	 \\ \vspace{0.1\columnwidth}
   	 \hfill\hfill \includegraphics[height=0.23\columnwidth]{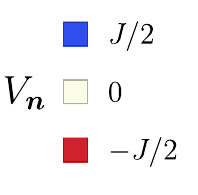}
   }   
   \subfloat[Example lattice]{\includegraphics[height=0.5\columnwidth]{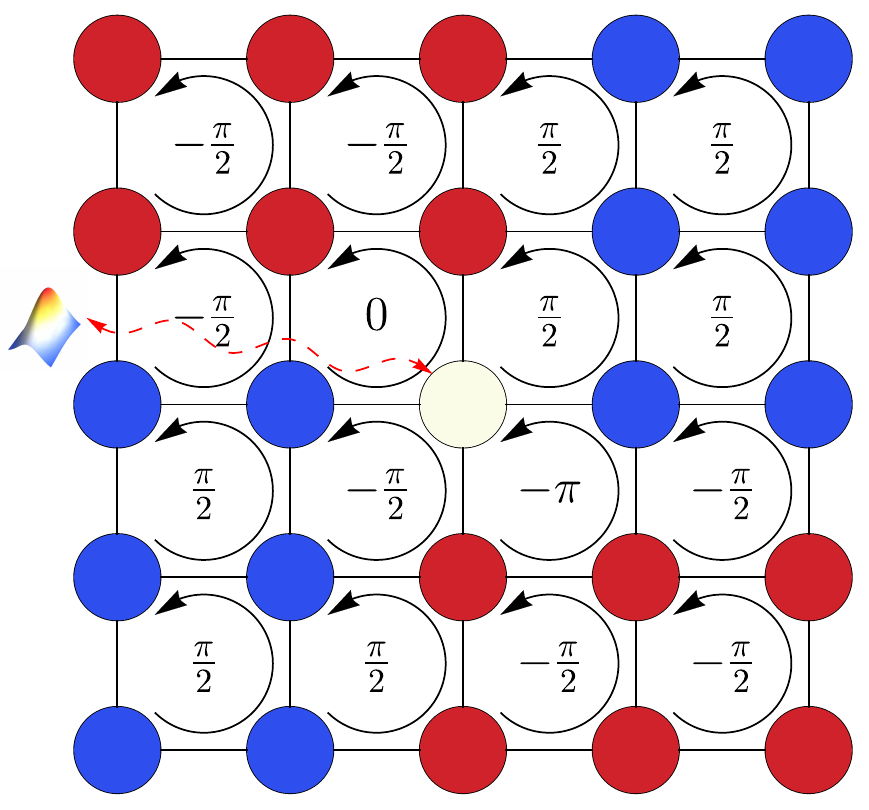}\label{fig:4foldlat}}
   \hfill\hfill

   \caption{
   \protect\subref{fig:4foldflux} Calculation of flux passing through a plaquette, as in \cref{eq:plaquette}: it is equal to the cumulative phase gained when hopping around the plaquette.
   \protect\subref{fig:4folddV} Robustness of the lattice to the appearance of dark modes. We test two on-site potential patterns: an saddlepoint-like form (solid blue line) and an alternating fixed potential (dashed green line). Varying the strength of potential, we numerically diagonalize $H^{\p{\rm 4f}}$ and plot its minimal mode wavefunction at the drain, $\lvert\psi^{\p{i}}_{\vn_{0}}\rvert^{2}$ and eigenenergy difference, $\abs{\gve_{i}-\gve_{j}}$. We look to maximize both, as far as possible: for the saddlepoint, this would be $\mathcal V\approx 3.5J$; for the fixed potential this is $\mathcal V \approx 0.5J$.
   \protect\subref{fig:4foldlat} Taking into account \cref{eq:4fconstraints} and the robustness analysis, we show one lattice realizing the symmetry of \cref{eq:s4fexplicitsym}. This is a two dimensional $5\times5$ array with uniform nearest-neighbor hopping strength $J$, on-site potential $V_{\vn} = \p{-1}^{q_{\vn}}J/2$ at each site outside the central drain, and flux through each plaquette as shown in the figure.
   }
   \label{fig:4fold}
\end{figure*}

One of the first correlation patterns observed in a chiral system was a ``rainbow'', or mirror, pattern, where the steady state takes the form of two-mode squeezing in real space, with correlations $\gs_{m,n} = \gd_{m,-n}$ \cite{Zippilli2015} . A similar form is seen in other one-dimensional systems \cite{Yanay2020}, as well as in the 2D Hofstadter lattice, depending on the placement of the drain \cite{Yanay2018}.
In these cases the symmetry matrix $\gs$ takes on a block-diagonal form, with blocks of size $2\times2$. A natural extension would be to larger blocks; we thus set out to generate a symmetry pattern with blocks of size $4\times4$. 
In analog to the one dimensional case, we envision a square lattice where each set of four sites positioned at the corners of a square centered at the origin are entangled only with each other. This is shown in \cref{fig:4foldcorr}.
A similar process, we note, could be followed to produce a triangular, hexagonal, or any other kind of n-fold symmetry.

We take the lattice to be a two-dimensional square array of size $N = \p{2L+1}\times \p{2L+1}$, with sites labeled $\vn = \mat{n_{x},n_{y}}$, $-L\le n_{x},n_{y}\le L$.
We then divide the lattice into quadrants, defined by
\begin{equation}\begin{gathered}
\forall \vn \ne \mat{0,0}:	q_{\vn} = \fopt{1  & n_{x} >0, n_{y}\ge0 \\ 2& n_{x} \le0, n_{y}>0   \\ 3& n_{x} <0, n_{y}\le0  \\ 4 & n_{x} \ge 0, n_{y}< 0. } 
\label{eq:nquad}
\end{gathered}\end{equation}

With this definition, we can write the correlation matrix in an explicit a block diagonal form, as
\begin{equation}
\gs^{\p{\rm 4f}}_{\vec m,\vec n } = \fopt{1 & \vec m = \vec n = \mat{0,0} \\ 
	u^{4\times4}_{q_{\vm},q_{\vn}}\br{\vm}  & \exists l\in \acom{1,2,3,4}:  \vm = R^{l}\cdot \vn \\ 0 & \rm{o/w},}
\label{eq:s4fexplicitsym}
\end{equation}
where $R$ is the rotation by $\pi/2$ matrix and $u^{4\times4}\br{\vm} = u^{4\times4}\br{R\cdot \vm}$ are a set of symmetric, unitary $4\times4$ matrices. The matrix $\gs^{\p{\rm 4f}}$ has the required chiral structure for a drain site at the origin, satisfying \cref{eq:chiralscond} with ${\vn_{0} = \mat{0,0}}$, and it describes the correlation structure shown in \cref{fig:4foldcorr}.

There are a number of matrices $\gs^{\p{4\times4}}$ that could produce useful entanglement resources. We choose here
\begin{equation}
u^{\p{4\times4}}\br{\p{x,y}} = {\frac{\p{-1}}{\sqrt{2}}}^{\abs{x}+\abs{y}}
	\mat{0 & 1 & 0 & -i \\ 1 & 0 & i & 0 \\ 0 & i & 0 & -1 \\ -i & 0 & -1 & 0},
\label{eq:sigma4x4}
\end{equation}
having added a parity factor which to further simplify later calculations.

We now select a subset of elements to use in the lattice. Taking quadratic Hamiltonian of \cref{eq:HS}, we limit it to the on-site potentials and nearest-neighbor hopping that are common in experimental realizations, setting
\begin{equation}
H^{\p{4f}}_{\vm,\vn} = \gd_{\vm,\vn}V_{\vn} - \gd_{\abs{\vm-\vn},1}J_{\vm,\vn}.
\end{equation}

Next, we apply the chiral constraints to find what set of terms would stabilize the correlation structure. 
By requiring $H^{\p{4f}}, \gs^{\p{4f}}$ satisfy \cref{eq:chiralreq}, we find a set of constraints for $V_{\vn}, J_{\vm,\vn}$:
\begin{subequations}\label{eq:4fconstraints}\begin{gather}
V_{\tiny \mat{0,0}}  = 0, \label{eq:conV0}
\\ \forall \vn \ne \mat{0,0} : \qquad \quad \quad V_{R\cdot \vn} = -V_{\vn},\qquad \label{eq:conVQ}
\\ \forall \vec m,\vec n \ne \mat{0,0} : \qquad J_{R\cdot \vec m,R\cdot \vec n} = i^{q_{\vm} - q_{\vn}}J_{\vec m,\vec n}^{*}.  \label{eq:conJQ}
\\ J_{\tiny \mat{0,0},\mat{\pm1,0}}  = \pm \tfrac{1}{\sqrt{2}}\p{J_{\tiny \mat{0,0},\mat{0,\pm1}}^{*} + i J_{\tiny \mat{0,0},\mat{0,\mp1}}^{*}}, \label{eq:conJQ0}
\end{gather}\end{subequations}
We see \cref{eq:conV0} requires that the drain has no on-site potential. \Cref{eq:conVQ,eq:conJQ} dictate the relation between the four quadrants, with on-site potentials flipping signs and hopping terms flipping their phase with every rotation. Finally, \cref{eq:conJQ0} constrains the coupling of the drain to the four quadrants.

\Cref{eq:conJQ0,eq:conJQ} describe the constraints in terms of the hopping parameters $J_{\vm,\vn}$. As these are gauge-dependent quantities, it is useful to restate the constraints in terms of a physical quantity, the flux threaded through each four-site plaquette. As sketched out in \cref{fig:4foldflux}, the flux is equal to the total phase gained by hopping around the loop. For the plaquette centered at $\vn + \mat{\half,& \half}$, it is given by
\begin{equation}\begin{split}
&\Phi \br{\vn + \mat{\half,\half}}= \arg\Big[ J_{\vn,\vn + {\tiny \mat{0,1}}}\times
		 \\ &\quad J_{\vn + {\tiny \mat{0,1}},\vn + {\tiny \mat{1,1}}} \times J_{\vn + {\tiny \mat{1,1}},\vn + {\tiny \mat{1,0}}} \times J_{\vn + {\tiny \mat{1,0}},\vn}\Big].
\label{eq:plaquette}
\end{split}\end{equation}
It follows from \cref{eq:conJQ} that for any plaquette outside the central four surrounding the drain, the flux is flipped in direction with every $\half[\pi]$ rotation,
\begin{equation}
\forall  \vec p \ne \mat{\pm\half,\pm \half} : \qquad \Phi\br{R\cdot \vec p} = -\Phi\br{\vec p}.
\end{equation}

At the center, this relation does not hold. Instead, we find that the total flux through these plaquettes is fixed,
\begin{equation}\begin{split}
\Phi\br{\mat{\half,\half}} + \Phi\br{\mat{-\half,\half}} + \qquad & 
\\ \Phi\br{\mat{-\half,-\half}} + \Phi\br{\mat{\half,-\half}} & = \pi.
\end{split}\end{equation}
Notably, this means realizing the four-fold entanglement lattice requires breaking of time-reversal symmetry, that is, the presence of phases.

Taking all of the above into account, the constraints of \cref{eq:4fconstraints} still allow us to freely choose the parameters of a single quadrant. As we have discussed, these can be chosen to satisfy experimental demands ensure that there are no dark modes. 
We demonstrate this process for a $5\times5$ lattice: we begin by uniformly setting all hopping elements to $\abs{J_{\vm,\vn}} = J$, and $\Phi\br{\vec p} = \half[\pi]$ in all plaquettes in the first quadrant. This leaves the on-site potential terms as the remaining variables. To choose those, we numerically evaluate a number of potential maps to see which is most robust to the appearance of dark modes, as shown in \cref{fig:4folddV}. We choose one to arrive at a lattice, sketched out in \cref{fig:4foldlat}, which stabilizes the correlations of \cref{eq:s4fexplicitsym}. 

 \begin{figure}[ttbp] 
   \centering
   \hfill
   \subfloat[Stabilization]{\includegraphics[width=0.3\columnwidth]{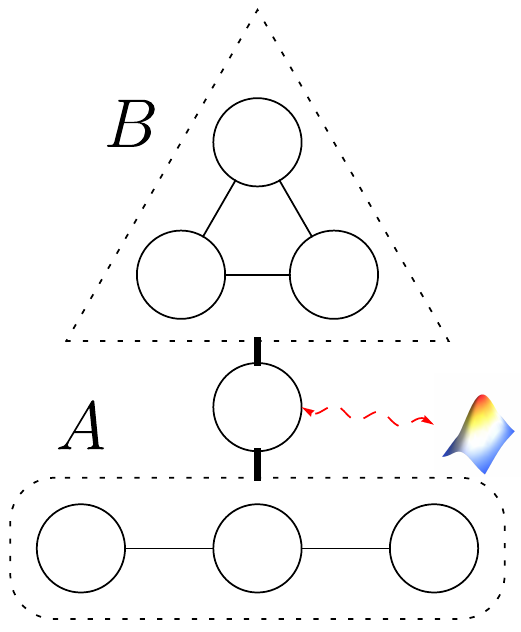} \label{fig:heraldlat}}\hfill
   \hfill
   \subfloat[Measurement]{\includegraphics[width=0.3\columnwidth]{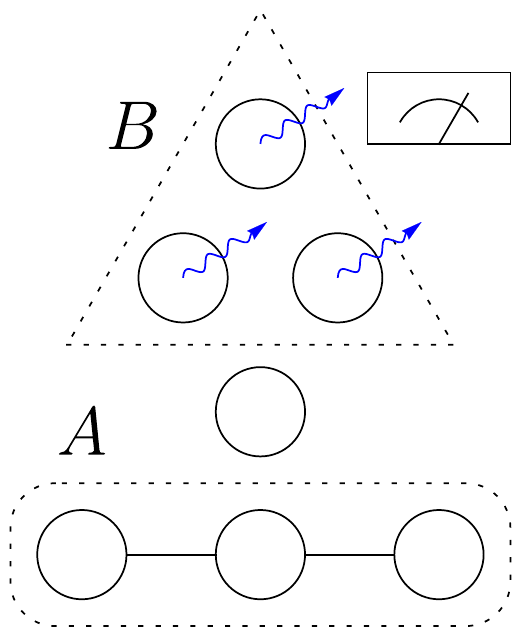} \label{fig:heraldmeas}}\hfill
   \hfill
   \subfloat[Final state]{\includegraphics[width=0.3\columnwidth]{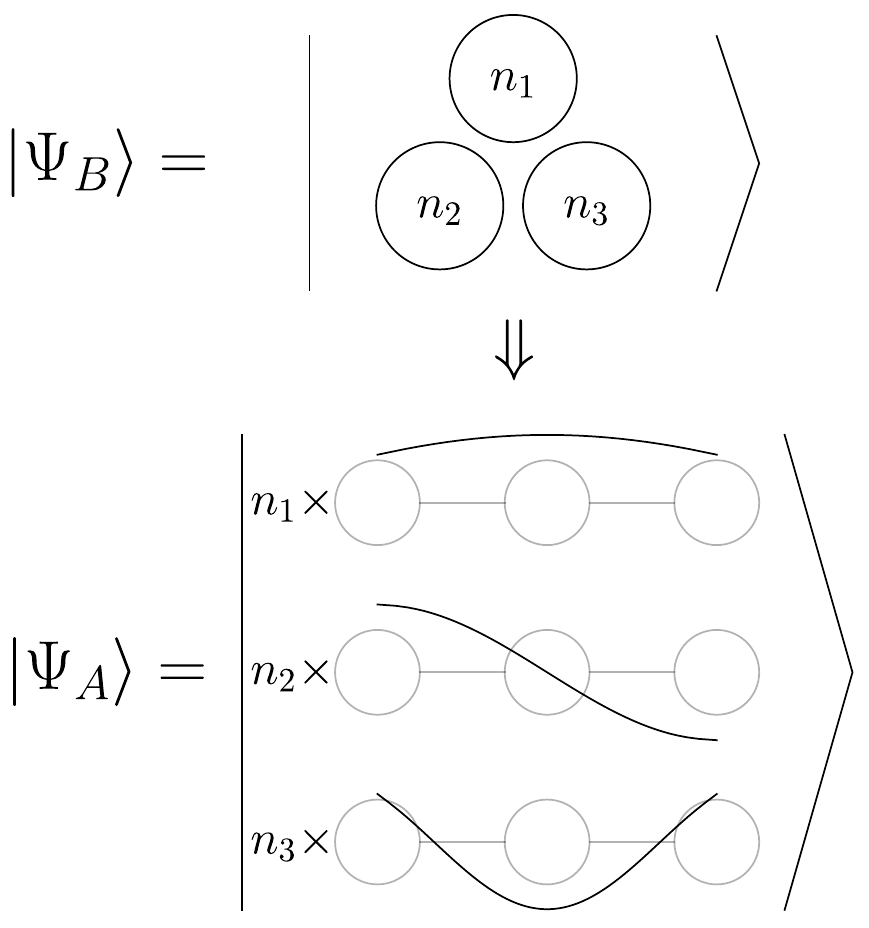} \label{fig:heraldout}}\hfill
   \hfill
   \caption{Use of a stabilized heralding state to prepare a known momentum eigenstate.
   \protect\subref{fig:heraldlat} The dissipative lattice includes the primary subsystem $A$, an auxiliary heralding subsystem $B$, and a drain site. It is prepared according to the chiral symmetry constraints and allowed to relax, stabilizing the squeezed state described by \cref{eq:heraldnkcorr}.
   \protect\subref{fig:heraldmeas} The drain site and all the couplings in the heralding lattice $B$ are disconncected, and a number measurement is performed on each site in $B$.
   \protect\subref{fig:heraldout} The measurement collapses $B$ into a Fock number state, and due to the prepared entangled state, collapses system $A$ into the same Fock state in its eigenmode basis.
   }
   \label{fig:heralding}
\end{figure}

\pardash{Momentum Eigenmode Heralding}

Our second example is motivated by heralding applications in quantum information. In a quantum computation device, herald photons are ones that are entangled with the main computational state in a way that their measurement allows for post-selection or post-processing of the results. It has been used, e.g.~as a replacement for non-linear elements in optics-based quantum computation \cite{Knill2001,OBrien2003,Pittman2003}, and in realizing a Boson sampling system which requires Fock-state inputs \cite{Aaronson2013,Tillmann2013}.

We set to design a system allowing the preparation of a known Fock state in some non-local basis. Consider a lattice partitioned into a single drain site, a primary lattice $A$ and a heralding lattice $B$. We set out to stabilize the Gaussian state
\begin{equation}
\ket{\psi_{\rm ss}} \propto \exp\bmat{e^{i\phi }\tanh r \sum_{k}\hat \ga_{k}\dg\hat b_{k}\dg}\vac
\label{eq:heraldnkcorr}
\end{equation}
where $\hat \ga_{k}$ is the annihilation operator for the $k$-th energy eigenmode of lattice $A$ and $\hat b_{k}$ is the annihilation operator for site $k$ of lattice $B$. 
In this product state each site in lattice $B$ is two-mode squeezed with an \emph{energy eigenmode} in lattice $A$, which is to say, the number of photons in modes $\hat\ga_{k}, \hat b_{k}$ are identical, 
\begin{equation}\begin{split}
\ket{\psi_{\rm ss}} = \prod_{\otimes k}\Big(\frac{1}{\cosh r}\sum_{n}\p{e^{i\phi }\tanh r}^{n} \ket{n}_{\ga_{k}}\ket{n}_{b_{k}}\Big).
\end{split}\end{equation}
After preparing the state, the systems are separated, and number measurements are made on system $B$. Due to the nature of two mode squeezing, this collapses the state of $A$ into a Fock state in its energy eigenmode basis, with the occupation of each mode corresponding to the measured result in the corresponding site in $B$. This procedure is outlined in \cref{fig:heralding}.

As an example, we make $A$ a one-dimensional chain of length $L$ with nearest-neighbor hopping. The stabilizing lattice then has $N=2L+1$ sites. We label the drain as $n=0$, the primary lattice by $1\le n\le L$ and the heralding lattice by $-L\le n\le -1$.
The eigenmodes above are then given by  ${\hat \ga_{k} = \sqrt{\tfrac{2}{L+1}}\sum_{n=1}^{L} \sin \tfrac{\pi kn}{L+1} \hat a_{n}}$, $\hat b_{k} = \hat a_{-n}$ for $1\le k\le L$.
Rewriting \cref{eq:heraldnkcorr} in terms of \cref{eq:GCSss}, we find the desired symmetry matrix
\begin{equation}
\gs^{\p{\rm h}}_{m,n} = \fopt{ 1 & m=n=0
\\ \sqrt{\tfrac{2}{L+1}}\sin{\tfrac{\pi m n}{L+1}}  & \sign\br{m}=-\sign\br{n}
\\ 0 & \text{o/w}. }
\end{equation}
It is easy to verify that $\gs^{\p{\rm h}}$ satisfies the chiral conditions of \cref{eq:chiralscond}.

We will now construct a system which stabilizes this state. We begin with the general Hamiltonian of \cref{eq:HS}. As we have chosen the sublattice $A$ to represent a one-dimensional chain with nearest-neighbor hopping, we have for the positive indices,
\begin{equation}
\forall m,n>0: \quad H^{\p{\rm h}}_{m,n} = V\gd_{m,n} - J \gd_{\abs{m-n},1},
\label{eq:HheraldA}
\end{equation}
for some hopping strength $J$ and overall energy offset $V$. 

Next, we require $H^{\p{\rm h}},\gs^{\p{\rm h}}$ satisfy \cref{eq:chiralreq}, and find the constraints on the remainder of the matrix
\begin{subequations}\label{eq:hconstraints}\begin{gather}
\label{eq:conherdrain}
	H_{0,0}  = 0 
\\ \nonumber \forall m,n>0 :
\\ \label{eq:conherHAHB}
	H_{-{m},-{n}} =  -\tfrac{2}{L+1}{\sum}_{j,l=1}^{L}H_{j,l}^{*}\sin\tfrac{\pi {m}j}{L+1}\sin\tfrac{\pi {n}l}{L+1},  
\\ \label{eq:conherAB}
	H_{-{m},{n}} =  -\tfrac{2}{L+1}{\sum}_{j,l=1}^{L}H_{j,-l}^{*} \sin\tfrac{\pi {m}j}{L+1}\sin\tfrac{\pi {n}l}{L+1}, 
\\ \label{eq:conherJAJB}
	H_{0,{n}} = \sqrt{\tfrac{2}{L+1}}{\sum}_{l=1}^{L}H_{0,-l}^{*}\sin\tfrac{\pi {n}l}{L+1}. 
\end{gather}\end{subequations}

As before, we find from \cref{eq:conherdrain} that the drain must be set at the center of the energy spectrum. 
\Cref{eq:conherHAHB} defines the dynamics of the heralding lattice. Substituting from \cref{eq:HheraldA}, we find
\begin{equation}
\forall m,n>0 : H_{-{m},-{n}} = \gd_{m,n}\p{2J \cos\br{\tfrac{\pi m}{L+1}} - V}.
\end{equation}
The sublattice $B$ is thus made up of disjoint sites, each with on-site potential equal to the energy of the eigenmode of $A$ that it is coupled to. This reflects the two-mode squeezing of energy eigenstates that characterizes this chiral symmetry \cite{Yanay2018}.

\begin{figure}[tbp] 
   \centering
	\includegraphics[width=\columnwidth]{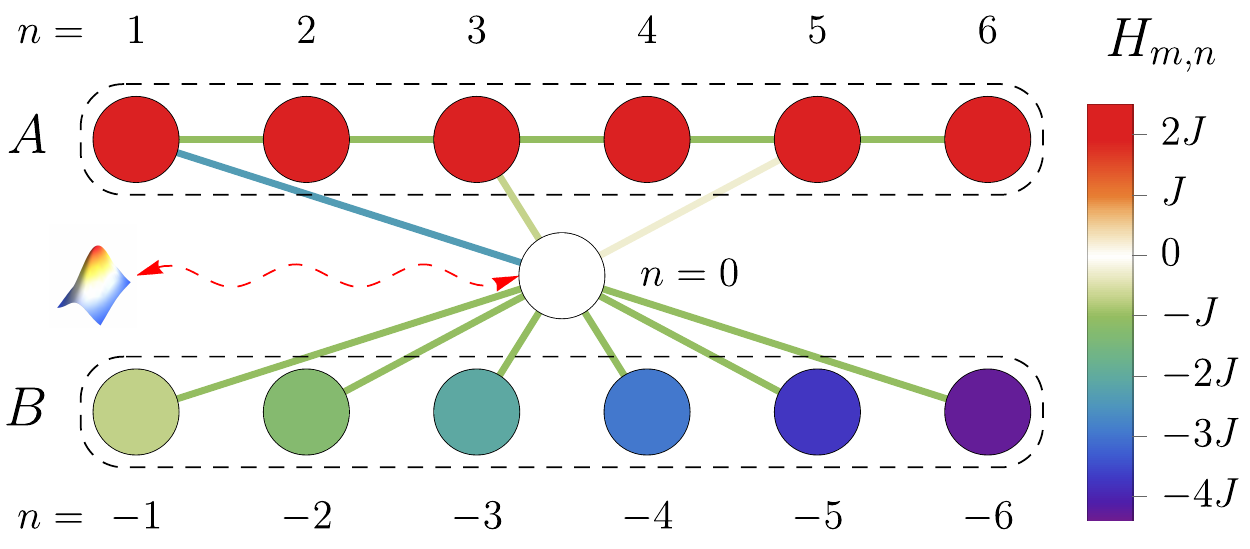}

   \caption{A sample system stabilizing a heralding state. Here, the color of each node and vertex corresponds to the on-site potential and hopping strength, respectively. In this example $\forall m,n>0,{H_{n,n+1} = H_{0,-n} = -J},H_{n,n} = 2.5J,H_{m,-n}=0$, and the rest of the lattice is dictated by \cref{eq:hconstraints}.
   At the steady state, each site in system $B$ is in a two-mode squeezed state with one momentum eigenmode of $A$, as in \cref{eq:heraldnkcorr}.
   }
   \label{fig:heraldingex}
\end{figure}

This leaves the inter-system and drain-system coupling terms. \Cref{eq:conherAB} is easily satisfied by setting ${H_{-{m},{n}} = 0}$, i.e.~by decoupling $A$ from $B$. The coupling from the drain to $A$ and $B$ is dictated by \cref{eq:conherJAJB}, and as before they can be chosen numerically to remove any dark states. One simple choice is a constant coupling to the sites of $B$, $H_{0,-{n}} = -J$. This version of the system is shown in \cref{fig:heraldingex}.

It is notable that the symmetry of the system, and indeed the steady state, are invariant under uniform scaling of all $H_{0,n}$ (or of all $H_{m,-n}$) This greatly simplifies the decoupling phase described in \cref{fig:heraldmeas}: as long as the coupling terms are reduced uniformly, they can be turned off slowly with without affecting the entangled state of the system.

\pardash{Outlook}
Both systems we have presented can be immediately implemented to produce useful entanglement resources. 
The lattice outlined in \cref{fig:4foldlat}, including the required fluxes, could be realized in a microwave cavity array \cite{Anderson2016,Owens2018}. 
Quadripartite mode entanglement has been produced within a single cavity in an optical frequency comb \cite{Pysher2011}; the method we propose here generates spatially separated entanglement in separate cavities, and uses a robust reservoir engineering technique.
The lattice shown in \cref{fig:heraldingex}, which requires no fluxes, is even easier to implement, and could be a resource for boson sampling calculations \cite{Knill2001,OBrien2003,Pittman2003,Aaronson2013,Tillmann2013}. 
Finally, the algorithm we have outlined is quite general and could be a powerful tool in preparing the sort of entangled states that are a critical of any quantum computing setup.

\pardash{Acknowledgements}
I would like thank A.~Clerk for useful discussions on chiral symmetry, and R.~Ruskov for proof-reading and giving insightful comments on the manuscript.

\end{document}